\documentclass{proc}
\usepackage{t1enc}
\usepackage{amssymb}

\usepackage{edbkps}

\usepackage[dvips]{graphicx}

\let\footnote\savefootnote
\let\footnotetext\savefootnotetext
\let\footnoterule\savefootnoterule

\normallatexbib %will produce bibliography entries as shown in the
\begin{document}

\articletitle
{A New approach to the Study\\ of the Smith + Smith Conjecture
}

\author{R. P. Mondaini, N. V. Oliveira}

\altaffiltext{}{Alberto Luiz Coimbra Institute\\
for Graduate Studies and Research in Enginerring,
COPPE/UFRJ - Technology Centre,\\ P.O. Box 68511, Rio de Janeiro, RJ,
Brazil,
\email{rpmondaini@gmail.com}}

\begin{abstract}
The search for a point set configurations of the $\mathbb{R}^3$ space
which contains the smallest value of the Euclidean Steiner Ratio is
almost finished. In the present work we introduce some analytical
methods which aim to support a famous conjecture of Discrete Mathematics
literature. The relations of this problem with that of molecular
architecture in terms of motivation as well as application of the
formulae obtained is also emphasized.
\end{abstract}

\begin{keywords}
Euclidean Full Steiner Trees, Steiner Ratio Function, Macromolecular structure modelling.
\end{keywords}

\section{Introduction}
One of motivations for modelling the macromolecular structure is
the careful observation of the protein data banks. In the present
stage of macromolecular evolution, the site of the atoms in a
biomacromolecule can be modelled by Steiner points of an usual Steiner
Problem. This is the Steiner Tree modelling with three edges  converging
at a Steiner Point and each pair of edges making angles of $120^{\circ}$ therein.
This looks as a last stage in the Nature's plans for modelling these
structures since it can be shown that this tree conformation is the most
stable one \cite{mondaini}. After an exhaustive series of computational
experiments by working with evenly spaced given points along a right
circular helix of unit radius or,
\begin{equation}\label{eq1}
P_{j}=(\cos(j\omega),\,\sin(j\omega),\,\alpha j \omega); \quad 0 \leq j \leq n-1
\end{equation}where $\omega$ is the angular distance between consecutive  points
and $2\pi\alpha$ is the helix pitch, we get a sequence of Steiner points in a helix of smallest radius and the same pitch
$2\pi\alpha$, {\it i. e.}, belonging to the same helicoidal surface. All the experimental
results led to write the formula for angular distance of consecutive Steiner points
along the internal helix as \cite{ochoa}
\begin{equation}\label{eq2}
\omega_{k}=\arctan \left(\frac{y_{k}}{x_{k}}\right)+2\pi  \left[\frac{z_{k}}{2\pi\alpha}\right]+ \pi \left[\frac{m}{2}\right], \quad 1\leq k \leq n-2
\end{equation} and $\omega_{k+1}-\omega_{k} \approx b\omega + c$, with $b\approx1$, $c\approx0$. The squared brackets
stand for the greatest integer value and $m=1, \,2,\,3,\,4$, according to the quadrant of the angle  $\arctan \left(\frac{y_{k}}{x_{k}}\right)$.

We have worked with a modified version of a famous algorithm \cite{smith}. This reduced search space version of the algorithm
was developed in 1996 and is available at  {\it http:\slash\slash www.biomat.org
\slash apollonius}.

From eq. (\ref{eq2}), we write the Ansatz for the Steiner Points $S_{k}(x_{k},y_{k},z_{k})$ as
\begin{equation}\label{eq4}
S_{k}(r(\omega,\alpha)\cos(k \omega),r(\omega,\alpha)\sin(k \omega), \alpha k \omega), \quad 1 \leq k \leq n-2.
\end{equation}

By forming a Steiner tree with points $P_{j}$ and $S_{k}$ with a 3-Sausage's topology \cite{smithsmith}, we can obtain
from the condition of edges intersecting at $120^{\circ}$, an expression for the radius function $r(\omega,\alpha)$ or:
\begin{equation}
r(\omega,\alpha)=\frac{\alpha \omega}{\sqrt{A_{1}(1+A_{1})}}
\end{equation} where \begin{equation}A_{1}=1-2\cos\omega. \end{equation}

The restriction to full Steiner trees can be obtained from eq. (\ref{eq1})
alone. We then have for the angle of consecutive edges formed by the
given points
\begin{equation}
-\frac{1}{2} \leq \cos \theta(\omega,\alpha)= -1 + \frac{(1+A_{1})^{2}}{2(\alpha^{2}\omega^{2}+1+A_{1})}.
\end{equation}

\section
{The generalization of the formulae to sequences of non-consecutive points}

The generalization of the formulae derived in section 1 to non-consecutive points \cite{arxiv}
along a right circular helix is done by taking into consideration the possibility of skipping
points systematically to define subsequences. Let us consider the subsequences of fixed and Steiner
points respectively.

\begin{equation}
(P_{j})_{m,\,l_{max}^{j}}:\, P_{j},\, P_{j+m},\, P_{j+2m},\,...,\,P_{j+l_{max}^{j}.\,m}
\end{equation}

\begin{equation}
(S_{k})_{m,\,l_{max}^{k}}:\, S_{k},\, S_{k+m},\, S_{k+2m},\,...,\,
S_{k+l_{max}^{k}.\,m}
\end{equation}\\with $0 \leq j \leq m-1 \leq n-1,  \quad   0 \leq k \leq m-1 \leq n-2$ and
\begin{equation}
l_{max}^{j}=\left[\frac{n-j-1}{m}\right]; \quad l_{max}^{k}=\left[\frac{n-k-2}{m}\right]
\end{equation}where $(m-1)$ is the number of skipped points and the square brackets stand for
the greatest integer value. It is worth to say that the sequences (\ref{eq1}) and (\ref{eq4})
correspond to $(P_{0})_{1,n-1}$ and $(S_{1})_{1,n-2}$ respectively.

The $n$ points of the helical point set are then grouped into $m$ subsequences and we
consider a new sequence of $n$ given points which is written as
\begin{equation}
P_{j}=\bigcup_{j=0}^{m-1}\left(P_{j} \right)_{m,\,l_{max}^{j}}
\end{equation}

Analogously, a new sequence for the union of the subsequences of the Steiner points
is introduced in the form
\begin{equation}
S_{k}=\bigcup_{k=1}^{m-1}\left(S_{k} \right)_{m,\,l_{max}^{k}}
\end{equation}

If the points $P_{j+lm},$  $S_{k+lm}$ are evenly spaced along the helices,
their coordinates are given analogously to eqs. (\ref{eq1}) and (\ref{eq4}) of section 1.
We can write,
\begin{equation}
P_{j+lm}\left(\cos(j+lm)\omega, \sin(j+lm)\omega, \alpha(j+lm)\omega  \right)
\end{equation}
\begin{equation}
S_{k+lm}\left(r_{m}(\omega,\alpha)\cos(k+lm)\omega, r_{m}(\omega,\alpha)\sin(k+lm)\omega, \alpha(k+lm)\omega  \right).
\end{equation}

We can now form Steiner trees with the points $P_{j+lm}$ and $S_{k+lm}$ with the 3-Sausage's
topology. We get from the condition of an angle of $120^{\circ}$ between each pair of edges
meeting at a Steiner point, the expression of radius $r_{m}(\omega,\alpha)$,
\begin{equation}
r_{m}(\omega,\alpha)=\frac{m\alpha\omega}{\sqrt{A_{m}(1+A_{m})}}
\end{equation} where \begin{equation} A_{m} = 1 - 2 \cos(m\omega). \end{equation}

The restriction to full Steiner trees is now obtained from the points $P_{j+lm}$ only and we have
\begin{equation}\label{eq16}
-\frac{1}{2} \leq \cos \theta_{m}(\omega,\alpha) = -1 + \frac{(1+A_{m})^2}{2(m^{2}\alpha^{2}\omega^{2}+1+A_{m})}
\end{equation}

\section
{A Proposal for a Steiner Ratio Function}

After a straightforward but tedious calculation, we get that the Euclidean lengths of the $m$-spanning tree
and of the $m$-Steiner tree for a great number of fixed points, $n>>1$, are given by
\begin{equation}
l_{SP}^{m}(\omega,\alpha)=n \sqrt{m^{2}\alpha^{2}\omega^{2}+1+A_{m}}
\end{equation}
\begin{equation}
l_{ST}^{m}=n\left(1+m\alpha\omega\sqrt{\frac{A_{m}}{1+A_{m}}}\right).
\end{equation}

The usual prescription for the Steiner Ratio Function (SRF) lead us to write
\begin{equation}\label{eq19}
\rho(\omega,\alpha)=\frac{\displaystyle{\min_{m}}(1+m\alpha\omega\sqrt{\frac{A_{m}}{1+A_{m}}})}{\displaystyle{\min_{m}} \sqrt{m^{2}\alpha^{2}\omega^{2}+1+A_{m}}}
\end{equation}

The ``$\displaystyle{\min_{m}}$'' in eq. (\ref{eq19}) should be understood in the sense of
a piecewise function formed by the functions corresponding to the values $m=1,\,2,\,...,\,n-1.$

The restriction to full trees  is applied by remembering that for
3-dimensional macromolecular structure there is not tree built from partial full trees. The
tree which represents the ``scaffold'' of the structure is itself a full tree or it is
completely degenerate. It is seen that the surfaces corresponding to eq. (\ref{eq16}) for $m \geq 2$
violate the inequality there for a large part of their domain. The surface for $m=1$ has the largest
feasible domain. Our proposal for the SRF function of a helical point set should be written then as
\begin{equation}\label{eq20}
\rho(\omega,\alpha)=\frac{1+\alpha\omega\sqrt{\frac{A_{1}}{1+A_{1}}}}{\displaystyle{\min_{m}}\,\,\sqrt{ m^{2}\alpha^{2}\omega^{2}+1+A_{m}}}.
\end{equation}

We can also suppose necessary bounds on $\rho(\omega,\alpha)$, or
\begin{equation}\label{eq21}
\frac{\sqrt{3}}{3} \leq \rho(\omega,\alpha) \leq 1
\end{equation}where the first inequality stands for the Graham-Hwangs's greatest lower bound \cite{graham} for the Euclidean Steiner
Ratio.

The corresponding $\omega$-region is given by
\begin{equation}\label{eq22}
\arccos(1/4) \leq \omega \leq 2\pi - \arccos(1/4).
\end{equation}

For the region defined by eqs. (\ref{eq21}) and (\ref{eq22}), we can define the unconstrained optimization problem
of eq. (\ref{eq20}) in the form
\begin{equation}\label{eq23}
\rho(\omega,\alpha)= \displaystyle{\max_{m}}\,\, \rho_{m}(\omega,\alpha)
\end{equation} where the surfaces $\rho_{m}(\omega,\alpha)$ are
\begin{equation}
\rho_{m}(\omega,\alpha)=\frac{1+\alpha\omega\sqrt{\frac{A_{1}}{1+A_{1}}}}{\sqrt{m^{2}\alpha^{2}\omega^{2}+1+A_{m}}}.
\end{equation}

The first three surfaces meet at a possible global minimum of the surface (\ref{eq20}) or (\ref{eq23}).
Furthermore, in this point, beyond the 3-Sausage's topology for the Steiner tree structure we
also have a 3-Sausage's configuration for the fixed points or the familiar set of vertices of regular
tetrahedra glued together at common faces \cite{toppur}. This non trivial solution can be written

\begin{equation}
\omega_{R}=\pi - \arccos(2/3); \quad \alpha_{R}=\sqrt{30}\left[9(\pi-\arccos (2/3))\right]
\end{equation} and

\begin{equation}\label{eq26}
\rho(\omega_{R},\alpha_{R})=\frac{1}{10}(3\sqrt{3}+\sqrt{7})=0.78419037337...
\end{equation} which is the value assumed by the main conjecture of authors of
ref. \cite{smithsmith} for the Steiner Ratio in $\mathbb{R}^{3}$ with Euclidean
distance.

It is worth to notice that there is always a pair of values which give the same
value of $\rho$, namely ($\omega,\alpha$) and $\left(2\pi N-\omega,\frac{\omega\alpha}{2\pi N -\omega}\right),\,\, N \in\, \mathbb{Z}.$ This
means that
$$\bar{\omega}=\pi + \arccos(2/3), \quad \bar{\alpha}={\sqrt{30}\left[9(\pi+\arccos(2/3))\right]}$$
correspond to the same $\rho$-value, eq. (\ref{eq26}).

\section
{The Existence of a Global Minimum. The Weierstrass Theorem}

The use of the Weierstrass Theorem has as a requirement the definition of a compact domain.
We shall define it in the diagram below where we have depicted the curves $\rho_{m}=1$, $m=1,2,3,4,5$.

\begin{figure}[ht]
\begin{center}
\includegraphics[scale=1.0]{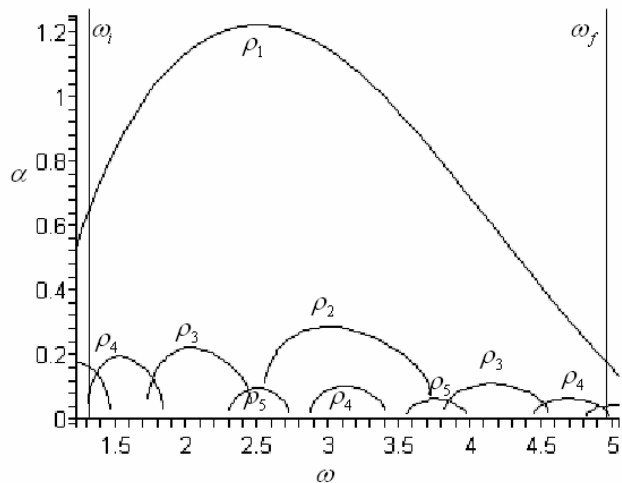}
\end{center}
\caption{The compact region to be used in the Weierstrass Theorem $\omega_{i}=\arccos(1/4)$,
$\omega_{f}=2\pi-\arccos(1/4)$ according the Graham-Hwang's lower bound.}
\end{figure}

The compact region can be defined as
\begin{equation}
(H\rho_{1}- \displaystyle{\bigcup_{k \geq \,2}} H\rho_{k}) \bigcap  \left\{(\omega,\alpha)| \arccos(1/4) \leq 2\pi - \arccos(1/4)\right\}
\end{equation} where $H\rho_{1}$, $H\rho_{k}$ are the hypographs of the functions $\rho_{1}=1$, $\rho_{k}=1$, respectively.
Since the functions $\rho_{k}(\omega,\alpha)$ are continuously decreasing from the boundaries of this compact
domain towards its interior, the Weierstrass Theorem guarantees the existence of a global minimum inside it.

\section
{Concluding Remarks}

This work is one of the last steps in our effort to proof the Smith+Smith Conjecture. There
are also nice results to be reported about the study of configurations in the neighbourhood of this value of the
Euclidean Steiner Ratio. They are associated to a continuous description of molecular chirality \cite{biomat3, chirality}
and their influence in the strengthening of molecular interaction. Studies of this subject are
now in progress and will be published elsewhere.

\end{document}